\documentclass[a4paper, 11 pt, twocolumn, final, conference]{IEEEtran}
\IEEEoverridecommandlockouts

\usepackage{graphicx}
\usepackage{times}   
\usepackage{amsmath} 
\usepackage{amssymb} 
\usepackage{amsfonts}
\usepackage{algorithm, algorithmic}
\usepackage{array}
\usepackage{multirow}
\usepackage{microtype}
\usepackage{graphics}
\usepackage{booktabs}
\usepackage{booktabs} 
\usepackage{url}

\usepackage{bm}
\usepackage{mystyle}
\usepackage{subcaption}

\makeatletter
\renewcommand*{\thetable}{\arabic{table}}
\renewcommand*{\thefigure}{\arabic{figure}}
\newcommand{\ra}[1]{\renewcommand{\arraystretch}{#1}}   
\let\c@table\c@figure
\makeatother 

\usepackage{caption}
\usepackage[noadjust]{cite}

\usepackage{todonotes}

\title{\LARGE \bf
Building Cross-Sectional Systematic Strategies By Learning to Rank
}

\author{Daniel Poh, Bryan Lim, Stefan Zohren, and Stephen Roberts
\thanks{D. Poh, B. Lim, S. Zohren and S. Roberts are with the Department of Engineering Science and the Oxford-Man Institute of Quantitative Finance, University of Oxford, Oxford, United Kingdom (email:\{dp, blim, zohren, sjrob\}@robots.ox.ac.uk).}
}

\begin{document}
\maketitle
\thispagestyle{empty}
\pagestyle{empty}
%
\begin{abstract}

The success of a cross-sectional systematic strategy depends critically on accurately ranking assets prior to portfolio construction. 
Contemporary techniques perform this ranking step either with simple heuristics or by sorting outputs from standard regression or classification models, which have been demonstrated to be sub-optimal for ranking in other domains (e.g. information retrieval).
To address this deficiency, we propose a framework to enhance cross-sectional portfolios by incorporating learning-to-rank algorithms, which lead to improvements of ranking accuracy by learning pairwise and listwise structures across instruments.
Using cross-sectional momentum as a demonstrative case study, we show that the use of modern machine learning ranking algorithms can substantially improve the trading performance of cross-sectional strategies -- providing approximately threefold boosting of Sharpe Ratios compared to traditional approaches.

\end{abstract}

\section{Introduction}\label{sec:intro}
Cross-sectional strategies are a popular style of systematic trading -- with numerous flavours documented in the academic literature across different trading insights and asset classes \cite{bazDissectingInvestmentStrategies2015}. In contrast to time-series approaches \cite{moskowitzTimeSeriesMomentum2012} which consider each asset independently, cross-sectional strategies capture risk premia by trading assets against each other -- buying assets with the highest expected returns and selling those with the lowest. The classical cross-sectional momentum strategy of \cite{jegadeeshReturnsBuyingWinners1993}, for instance, selects stocks by ranking their respective returns over the past year and betting that the order of returns will persist into the future -- buying assets in the top decile while selling the bottom decile. By trading winners against losers, cross-sectional strategies have been shown to be more insulated against common market moves, and perform in cases even when assets have non-negligible correlations (e.g. equity markets) \cite{bazDissectingInvestmentStrategies2015,amundiKeepUpMomentum,amundiUnderstandingMomentum}.

With the rise of machine learning in recent years, many cross-sectional systematic strategies which incorporate advanced prediction models have been proposed \cite{kimEnhancingMomentumStrategy2019, guEmpiricalAssetPricing2018, guAutoencoderAssetPricing2019}, 
often demonstrating significant improvements over traditional baselines. In general, these machine learning models are trained in a supervised fashion, and aim to minimise the mean squared error (MSE) of forecast returns over the holding time period. While the regression models under this framework accurately provide a mean estimate of future asset returns, they do not explicitly consider the \textit{expected ordering} of returns -- which is at the core of the design of cross-sectional strategies. 
This could hence have a negative impact on strategy performance, and consequently lead to sub-optimal investment decisions.

The limits of standard regression methods for ranking have been extensively studied in information retrieval applications, with a number of metrics and methodologies proposed \cite{pasumarthiTFRankingScalableTensorFlow2019}.
Collectively referred to as Learning to Rank (LTR) \cite{liLearningRankInformation2011}, today's variants make use of modern learning techniques such as deep neural networks and tree-based methods \cite{wangAttentionBasedDeepNet2017, liCombiningDecisionTrees2019, wuAdaptingBoostingInformation2010} -- leading to dramatic improvements in accuracy over simpler baselines.
 While LTR algorithms have also been used to a small degree in finance, the majority of papers focus on its use in simple stock recommendations \cite{WangMultiObjRankNets2019}, and lack a framework for the development and evaluation of general cross-sectional strategies.

In this paper, we show how LTR models can be used to enhance traditional cross-sectional systematic strategies -- adopting cross-sectional momentum as a demonstrative use-case. We first start by casting stock selection as a general ranking problem, which allows for the flexibility of switching between different LTR algorithms and incorporating state-of-the-art models developed in other domains. Next, we concretely evaluate LTR models against a mixture of standard supervised learning methods and heuristic benchmarks. Through tests on US equities, we demonstrate that the pairwise and listwise models better substantively improve the ranking accuracy of stocks -- leading to overall improvements in strategy performance with the adoption of LTR methods. Finally, although our ranking algorithms make use of momentum predictors, their modular nature allows these inputs to be adapted to incorporate other feature sets -- providing a novel and generalisable framework for general cross-sectional strategies.

\section{Related Works}

\subsection{Cross-Sectional Momentum Strategies}\label{sec:rw_csm}

Momentum strategies can either be categorised as (univariate) time series or (multivariate) cross-sectional. In time series momentum, an asset's trading rule depends only on its own historical returns.
It was first proposed by \cite{moskowitzTimeSeriesMomentum2012}, who documented the profitability of the strategy in trading nearly 60 different liquid instruments individually over 25 years. 
This has prompted numerous subsequent works such as \cite{bazDissectingInvestmentStrategies2015, rohrbachMomentumTraditionalCryptocurrencies2017, limEnhancingTimeSeries2019} which explore various trading rules alongside different trend estimation and position sizing techniques that are aimed at refining the overall strategy. 

Cross-sectional momentum employs a similar idea but focuses on comparing the relative performance of assets. It is characterised by first sorting the instruments by performance (typically taken to be returns), and then buying some fraction of top performers (winners) while simultaneously selling a similar-sized fraction of under-performers (losers). 
Since the earlier work in \cite{jegadeeshReturnsBuyingWinners1993} which considers this strategy for US equity markets, the literature has been replete with a spectrum of works -- ranging from those that report the existence of the momentum phenomenon in other markets and asset classes \cite{rouwenhorstInternationalMomentumStrategies1998, griffinMomentumInvestingBusiness2003, chuiIndividualismMomentumWorld2010, grootCrosssectionStockReturns2012, erbStrategicTacticalValue2006, lebaronTechnicalTradingRule1996}, to others that propose new ways to improve various aspects of the strategy. For instance, \cite{kimEnhancingMomentumStrategy2019} performs ranking with the forward-looking Sharpe ratio instead of historical returns. \cite{pirrongMomentumFuturesMarkets2005} constructs rankings based on returns standardised by their respective volatilities, arguing that this allows for a fairer comparison of instrument performance; \cite{bazDissectingInvestmentStrategies2015} employs a similar but more sophisticated approach by using volatility-normalised moving-average convergence divergence (MACD) indicators as inputs. A common thread across recent works is their use of a regress-then-rank approach \cite{wangStockRankingMarket2018} -- minimising  return predictions against targets before ranking the outputs and finally constructing a zero-investment portfolio by trading the tails of the sorted results. Notably, the loss commonly employed is the MSE (mean-squared error) which is a pointwise function, and some examples of works training with the MSE are  \cite{kimEnhancingMomentumStrategy2019} and \cite{guAutoencoderAssetPricing2019}. 

Surveying works related to cross-sectional momentum, we believe that this is the first paper to consider the use of ranking algorithms to enhance cross-sectional momentum strategies. Instead of ranking based on heuristics or on outputs produced by models trained on pointwise losses, we
propose using Learning to Rank algorithms -- demonstrating that learning the pairwise and listwise structure across securities produces better ranking and consequently better out-of-sample strategy performance.

\subsection{Learning to Rank in Finance}\label{sec:ltr_fin}

Learning to Rank is a key area of research in Information Retrieval and focuses on using machine learning techniques to train models to perform ranking tasks \cite{liLearningRankInformation2011, liuLearningRankInformation2011}. The information explosion along with modern computing advances on both the hardware (cloud-based GPUs and TPUs \cite{googlecloudCloudTPUDocumentation}) and software (open source Deep Learning frameworks such as {\tt Tensorflow} \cite{abadiTensorFlowLargescaleMachine2015} and {\tt PyTorch} \cite{paszkePyTorchImperativeStyle2019}) fronts have induced a shift in how machine learning algorithms are designed -- going from models that required handcrafting and explicit design choices towards those that employ neural networks to learn in a data-driven manner. This has prompted a parallel trend in the space of ranking models, with researchers migrating from using probabilistic varieties like the BM25 and LMIR (Language Model for Information Retrieval) that required no training \cite{liLearningRankInformation2011}, to sophisticated architectures that are built on the aforementioned developments \cite{pasumarthiTFRankingScalableTensorFlow2019}. 

Today, LTR algorithms play a key role in a myriad of commercial applications such as search engines \cite{liuLearningRankInformation2011}, e-commerce \cite{santuApplicationLearningRank2017} and entertainment \cite{pereiraOnlineLearningRank2019}. These algorithms have been explored to a more limited degree in finance, with specific applications in equity sentiment analysis and ranking considered in \cite{songStockPortfolioSelection2017, wangStockRankingMarket2018}.
\cite{songStockPortfolioSelection2017} apply RankNet and ListNet to ten years of market and news sentiment data, demonstrating higher risk-adjusted profitability over the S\&P 500 index return, the HFRI Equity Market Neutral Index (HFRI EMN) as well as pure sentiment-based techniques. On the other hand, \cite{wangStockRankingMarket2018} documents the superiority of LambdaMART over standard neural networks in predicting intraday returns on the Shenzhen equity exchange, adopting order book based features.

Despite the initial promising results, we note that these applications do not consider comparisons to traditional styles of systematic trading -- with comparisons only performed against market returns in \cite{songStockPortfolioSelection2017} and neural network models in \cite{wangStockRankingMarket2018} -- making it difficult to evaluate the true value added by LTR methods. The ranking methodologies are also customised to specific sentiment trading applications, making generalisation to other types of cross-sectional trading strategies challenging. We address these limitation explicitly in our paper -- proposing a general framework for incorporating LTR models in cross-sectional strategies, and evaluating their performance.

\section{Problem Definition}

Given a securities portfolio that is rebalanced monthly, the returns for a cross-sectional momentum (CSM) strategy at $\tau_m$ can be expressed as follows:
\begin{align}
r^{CSM}_{\tau_m,\tau_{m+1}} = \frac{1}{n_{\tau_m}} \sum^{n_{\tau_m} }_{i=1} X^{(i)}_{\tau_m} \frac{\sigma_{tgt}}{\sigma^{(i)}_{\tau_m}} r^{(i)}_{\tau_m,\tau_{m+1}}
\label{eqn:csm_rets} 
\end{align}
where $\tau_m, \tau_{m+1}\in \bm{\mathcal{T}} \subset \{1, ..., t-1, t, t+1, ... , T\}$ and $\bm{\mathcal{T}}$ denotes the set of indices coinciding with the last trading day of every month; $r^{CSM}_{\tau_m,\tau_{m+1}}$ are the realised portfolio returns from month $\tau_m$ to $\tau_{m+1}$, $n_{\tau_m}$ refers to the number of stocks in the portfolio and $X^{(i)}_{\tau_m} \in \{-1, 0, 1\}$ characterises the cross-sectional momentum signal or trading rule for security $i$.
We rebalance monthly to avoid excessive transaction costs that comes with trading at higher frequencies, daily for example.
We also fix the annualised target volatility $\sigma_{tgt}$ at 15\% and scale asset returns with $\sigma^{(i)}_{\tau_m}$ which is an estimator for ex-ante monthly volatility. In this paper, we use a rolling exponentially weighted standard deviation with a 63-day span on daily returns for $\sigma^{(i)}_{\tau_m}$, but note that more sophisticated methods (e.g. GARCH \cite{BollerslevGARCH}) can be used.

\subsection{Strategy Framework} \label{sec:csm_framework} 
The general framework for the CSM strategy comprises of the following 4 components.\\

\subsubsection{Score Calculation} 
\begin{equation}
    Y^{(i)}_{\tau_m} = f(\bm{u}^{(i)}_{\tau_m}),
\end{equation}
Presented with an input vector $\bm{u}^{(i)}_{\tau_m}$ for asset $i$ at $\tau_m$, the strategy's prediction model $f$ computes its corresponding score $Y^{(i)}_{\tau_m}$. For a cross-sectional universe of size $N_{\tau_m}$ at $\tau_m$, the list of scores of assets considered for trading is represented by the vector $Y_{\tau_m} = \{ Y^{(1)}_{\tau_m}, ... , Y^{(N_{\tau_m})}_{\tau_m}\}$.\\

\subsubsection{Score Ranking}
\begin{equation}
    Z^{(i)}_{\tau_m}=\mathcal{R}(Y_{\tau_m})^{(i)}
\end{equation}
with $Z^{(i)}_{\tau_m} \in \{1, ... , N_{\tau_m} \}$ being the position index for asset $i$ after applying the operator $\mathcal{R}(\cdot)$ to sort scores in ascending order.\\

\subsubsection{Security Selection}
\begin{align} 
    X^{(i)}_{\tau_m} = &
  \begin{cases}
  -1  & Z^{(i)}_{\tau_m} \leq \lfloor 0.1 \times N_{\tau_m} \rfloor \\
  1  & Z^{(i)}_{\tau_m} > \lfloor 0.9 \times N_{\tau_m} \rfloor \label{eqn:thresholding}\\
  0   & \text{Otherwise} \\
  \end{cases}
\end{align}
Selection is usually a thresholding step where some fraction of assets are retained to form the respective long/short portfolios. Equation \eqref{eqn:thresholding} assumes that we are using the typical decile-sized portfolios for the strategy (i.e. top and bottom $10\%$).\\

\subsubsection{Portfolio Construction}
Finally, simple portfolios can then be constructed by volatility scaling the selected instruments based on Equation \eqref{eqn:csm_rets}. 

In the following section, we provide an overview of score calculation techniques for both current strategy approaches and LTR models.

\section{Score Calculation Methodologies}

Most CSM strategies adhere to this framework and are generally similar over the last three steps, i.e., how they go about ranking scores, selecting assets and constructing the portfolio. However, they are particularly diverse in their choice of the prediction model $f$ used to calculate scores, ranging from simple heuristics \cite{jegadeeshReturnsBuyingWinners1993} to sophisticated architectures on an expansive list of macroeconomic inputs \cite{guEmpiricalAssetPricing2018}. While numerous techniques to compute scores exist, they can be grouped into 3 categories: Classical Momentum, Regress-then-Rank -- both of which are current approaches -- and finally Learning to Rank, which is our proposed method.

\subsection{Classical Cross-Sectional Momentum} \label{sec:srm_classical}

Classical variants of the CSM tend to lean towards the use of comparatively simple procedures for the score calculation. 

\vspace{10pt}
\textit{Jegadeesh \& Titman, 1993 \cite{jegadeeshReturnsBuyingWinners1993}:}

The authors who first documented the CSM strategy, propose scoring an asset with its raw cumulative returns, computed over the past 3 to 12 months:
\begin{align}
& \textrm{\textbf{Score Calculation:}}      & Y^{(i)}_{\tau_m} & = r^{(i)}_{\tau_m-252,\tau_m}
\end{align}
where $r^{(i)}_{{\tau_m}-252,\tau_m}$ is the raw returns over the previous 252 days (12 months) from $\tau_m$ for asset $i$. 

\vspace{10pt}
\textit{Baz et al., 2015 \cite{bazDissectingInvestmentStrategies2015}:}

A sophisticated alternative uses volatility-normalised MACD indicators as an intermediate signal, forming the final signal by combining indicators computed over different time scales. The indicator is given as:
\begin{align}
& \tilde{Y}^{(i)}_{\tau_m} = \xi^{(i)}_{\tau_m} / \textrm{std}(z^{(i)}_{\tau_m-252:\tau_m}) \label{eqn:baz_trading_sig}\\
& \xi^{(i)}_{\tau_m} = \textrm{MACD}(i, \tau_m, S, L) / \textrm{std}(p^{(i)}_{\tau_m-63:\tau_m}) \\
& \textrm{MACD}(i, \tau_m, S, L) = m(i, S) - m(i, L)
\end{align}
where $\textrm{std}(p^{(i)}_{\tau_m-63:\tau_m})$ represents the 63-day rolling standard deviation of security $i$, while $m(i, S)$ is an exponentially weighted moving average of prices for asset $i$ and $S$ translates to a half-life decay factor $HL = \log(0.5)/\log(1-1/S)$. The final composite signal combines different volatility-scaled MACDs over different time scales involving a response function $\phi(\cdot)$ and a set of short and long time scales $S_k \in \{8, 16, 32\}$ and $L_k \in \{24, 48, 96\}$ as set out in \cite{bazDissectingInvestmentStrategies2015}:
\begin{align}
    & \textrm{\textbf{Score Calculation:}} & Y_{\tau_m}^{(i)}=\sum^3_{k=1} \phi \bigl ( \tilde{Y}^{(i)}_{\tau_m}(S_k, L_k) \bigr ).\label{eqn:baz_fin_sig}
\end{align}

\subsection{Regress-then-Rank Method} \label{sec:srm_extensions}

Newer works employing a regress-then-rank approach typically compute the score via a standard regression (refer to Section \ref{sec:rw_csm}):

\begin{align}
& \textrm{\textbf{Score Calculation:}}      & Y^{(i)}_{\tau_m} & = f(\bm{u}^{(i)}_{\tau_m}; \bm \theta) \label{eqn:ml_pred}
\end{align}
where $f$ characterises a machine learning prediction model parameterised by $\bm \theta$ presented with some input vector $\bm{u}^{(i)}_{\tau_m}$. Using the volatility normalised returns as the target, the model is trained by minimising the loss, which is typically the MSE:
\begin{align}
    \mathcal{L(\bm\theta)} = & \frac{1}{M}\sum_\Omega \biggl ( Y^{(i)}_{\tau_m} - \frac{r^{(i)}_{\tau_m,\tau_{m+1}}}{\sigma^{(i)}_{\tau_m}} \biggr )^2 \label{eqn:ml_mse} \\
    \Omega = & \left \{ (Y^{(1)}_{\tau_1}, r^{(1)}_{\tau_1,\tau_2}/\sigma^{(1)}_{\tau_1}) , ..., \right . \nonumber \\  
    & \left . (Y^{(N_{\tau_{m-1}})}_{\tau_{m-1}}, r^{(N_{\tau_{m-1}})}_{\tau_{m-1},\tau_m} /\sigma^{(N_{\tau_{m-1}})}_{\tau_{m-1}}) \right \} \nonumber
\end{align}
where $\Omega$ represents the set of all $M$ possible forecasted and target tuples over the set of instruments and relevant time steps. 

\subsection{Learning to Rank Algorithms} \label{sec:srm_rankers}

LTR methods can be categorised as being pointwise, pairwise or listwise. The pointwise (pairwise) approach casts the ranking problem as a classification, regression or ordinal classification of individual (pairs of) samples, while the listwise approach on the other hand learns the appropriate ranking model by using ranking lists as inputs. 
In terms of ranking performance, the pointwise method has been observed to be inferior relative to the last two techniques \cite{liLearningRankInformation2011}.
Additionally, the loss function is not just the key difference across these models \cite{liLearningRankInformation2011} -- by incorporating the pairwise and listwise information across assets, it makes LTR models collectively distinct from the those outlined in Sections \ref{sec:srm_classical} and \ref{sec:srm_extensions}. 

We provide a high level overview of 4 LTR algorithms which we use in conjunction with the momentum strategy, highlighting the loss function but omitting technical details to keep our exposition brief.
For details adapting the LTR framework for the momentum strategy, please refer to Section \ref{sec:app_ltr4csm} in the Appendix.

\vspace{10pt}
\textit{Burges et al., 2005 (RankNet):}

While techniques that apply neural networks to the ranking problem already exist, RankNet was the first to train network based on \textit{pairs} of samples. Similar to contemporary methods, RankNet uses a neural network. Instead of minimising the MSE however, RankNet focuses on minimising the cross entropy error from classifying sample pairs -- optimising the network based on the probability that one element has a higher rank than the other. As training is conducted on individual pairs using stochastic gradient descent, RankNet has a complexity that scales quadratically with the number of securities at rebalance time.

\vspace{10pt}
\textit{Burges et al., 2010 (LambdaMART):}

LambdaMART is a state-of-the-art \cite{nguyenFactorizingLambdaMARTCold2016, bruchAlternativeCrossEntropy2020} pairwise method that combines LambdaRank \cite{burgesLearningRankNonsmooth2006} with Multiple Additive Regression Trees (MART). 
Interestingly, training in LambdaMART via LambdaRank does not involve directly optimising a loss function but rather making use of heuristic approximations of the gradients (referred to as $\lambda$-gradients) -- exploiting the fact that only the gradients and not actual loss values are required to train neural networks. 
This allows the models to circumvent dealing with the often flat, discontinuous and non-differentiable losses such as the NDCG (Normalised Discounted Cumulative Gain) \cite{jarvelinIREvaluationMethods2000}, which is simultaneously a popular position-sensitive information retrieval metric and that LambdaRank has been shown to locally optimise \cite{donmezLocalOptimalityLambdaRank2009, yueUsingSimultaneousPerturbation2007}. Given the formulation of $\lambda$-gradients, the loss involves the product of a pairwise cross entropy loss and the gain on some information retrieval metric (typically taken to be NDCG) \cite{wuAdaptingBoostingInformation2010}. MART on the other hand is a tree boosting method known for its flexibility. It also offer a simple way to trade off speed and accuracy via truncation which are important for time-critical applications such as search engines \cite{wuAdaptingBoostingInformation2010}.
LambdaMART which is the result of marrying these methods thus combines LambdaRank's observed empirical optimality (with respect to NDCG) \cite{donmezLocalOptimalityLambdaRank2009} with the flexibility and robustness of MART.

\vspace{10pt}
\textit{Cao et al., 2007 (ListNet):}

ListNet was developed to address the practical issues inherent with pairwise techniques such as their prohibitive computational costs, as well as their mismatched objective of minimising errors related to pairs classification instead of the overall ranking itself. ListNet resolves these problems by making use of a listwise loss, adopting a probablistic approach on permutations. By first computing ``top one" probability distributions over a list of scores and ground truth labels and then normalising each with a softmax operator, the loss is then defined to be the cross entropy between both these distributions. 
By using the entire cross-section of securities as inputs, ListNet has a complexity of $O(N_{\tau_m})$ at ${\tau_m}$ -- making it more efficient than RankNet which has a quadratic complexity of $O(N_{\tau_m}^2)$ since training is conducted on pairs. 

\vspace{10pt}
\textit{Xia et al., 2008 (ListMLE):}

Seeking to analyse and provide more theoretical support linking the choice of a ranking model's listwise loss function to its corresponding performance, \cite{xiaListwiseApproachLearning2008} proposed the use of the likelihood loss due to its nice properties of consistency, soundness and linear complexity. Additionally, the likelihood loss is continuous, differentiable and convex \cite{BoydConvexOptimization2004}. This culminated in the development of ListMLE -- a probabilistic ranking approach that casts the ranking problem as minimising the likelihood loss, or equivalently as maximising the likelihood function of a probability model. ListMLE has been shown to outperform other listwise methods on benchmark data sets \cite{xiaListwiseApproachLearning2008}, and shares the same linear complexity as ListNet. Given our results which we further discuss later in the paper, we note that the benefit of the linear complexity possessed by both ListMLE and ListNet might be relevant for larger data sets.

\subsection{Training Details}

RankNet, ListNet and ListMLE were trained with the {\tt Adam} optimiser based on each model's respective ranking loss function. Backpropagation was conducted for a maximum of 100 epochs where for a given training set, we partition 90\% of the data for training and leave the remaining 10\% for validation. As a matter of practicality, we set our target to be the returns 21 days ahead instead of the next month for training and validation. For Learning to Rank models using neural networks (i.e. RankNet, ListNet and ListMLE), we use 2 hidden layers but treated the width as a tunable hyperparameter. Early stopping was used to prevent model overfitting and is triggered when the model's loss on the validation set does not improve for 25 consecutive epochs. We also used dropout regularisation \cite{srivastavaDropoutSimpleWay2014} in the networks-based models as an additional safeguard against overfitting and similarly treated the dropout rates as a hyperparameter to be calibrated over model learning. Across all models, hyperparameters were tuned by running 50 iterations of search using {\tt HyperOpt} \cite{bergstraHyperoptPythonLibrary2015}. Further details on calibrating the hyperparameters can be found in Section \ref{app:training_details} of the Appendix.

\section{Performance Evaluation}

\subsection{Dataset Overview} \label{sec:pe_data_overview}

We construct our monthly portfolios using data from CRSP (Center for Research in Security Prices) \cite{wrds_crsp}. Our universe comprises of actively traded firms on NYSE from 1980 to 2019 with a CRSP share code of 10 and 11. At each rebalancing interval, we only use stocks that are trading above \$1. Additionally, we only consider stocks with valid prices and that have been actively trading over the previous year. All prices are closing prices.

\subsection{Backtest and Predictor Description} \label{sec:pe_backtest}

With the exception of both the classical strategies employing heuristic rankings, all other models were re-tuned at 5-year intervals. The weights and hyperparameters of the calibrated models were then fixed and used for out-of-sample portfolio rebalancing for the following 5-year window. The rebalancing takes place on the last trading day of each month. Focusing on ranking, we trade 100 stocks for each long and short portfolios at all times -- amounting to approximately $10\%$ of all tradeable stocks at each rebalancing interval. 
For predictors, we use a simple combination of the predictors employed by the classical approaches in Section \ref{sec:srm_classical}:
\begin{enumerate}
    \item \textit{Raw cumulative returns} -- Returns as per \cite{jegadeeshReturnsBuyingWinners1993} over the past 3, 6 and 12-month periods.
    \item \textit{Normalised returns} -- Returns over the past 3, 6 and 12-month periods standardised by daily volatility and then scaled to the appropriate time scale.
    \item \textit{MACD-based indicators} -- Retaining the final signal as defined in Equation \eqref{eqn:baz_fin_sig} from \cite{bazDissectingInvestmentStrategies2015}, we also augment our set of predictors by including the set of raw intermediate signals $\tilde{Y}^{(i)}_{\tau_m}(S_k, L_k)$ in Equation \eqref{eqn:baz_trading_sig} for $k=1,2,3$ computed at $t$ as well as for the past 1, 3, 6 and 12-month periods -- giving us a total of 16 features for this group.
\end{enumerate}

\subsection{Models and Comparison Metrics}
The LTR and reference benchmarks models (with their corresponding shorthand in parentheses) studied in this paper are:
\begin{enumerate}
    \item \textit{Random (Rand)} -- This model select stocks at random, and is included to provide an absolute baseline sense of what the ranking measures might look like assuming portfolios are composed in such a manner.
    \item \textit{Raw returns (JT)} -- Heuristics-based ranking technique based on \cite{jegadeeshReturnsBuyingWinners1993}, which is one of the earliest works documenting the CSM strategy.
    \item \textit{Volatility Normalised MACD (Baz)} -- Heuristics-based ranking technique with a relatively sophisticated trend estimator proposed by \cite{bazDissectingInvestmentStrategies2015}.
    \item \textit{Multi-Layer Perceptron (MLP)} -- This model characterises the typical Regress-then-rank techniques used by contemporary methods.
    \item \textit{RankNet (RNet)} -- Pairwise LTR model by \cite{burgesLearningRankUsing2005}.
    \item \textit{LambdaMART (LM)} -- Pairwise LTR model by \cite{burgesRankNetLambdaRankLambdaMART2010}.
    \item \textit{ListNet (LNet)} -- Listwise LTR model by \cite{caoLearningRankPairwise2007}.
    \item \textit{ListMLE (LMLE)} -- Listwise LTR model by \cite{xiaListwiseApproachLearning2008}.
\end{enumerate}

Performance of the various algorithms are finally evaluated with two sets of metrics -- the first involving those commonly found in finance (1 to 3), and the latter from the  information retrieval and ranking literature (4):
\begin{enumerate}
    \item Profitability: Expected returns ($\mathbb{E}[\textrm{Returns}]$) as well as the percentage of positive returns at the portfolio-level obtained over the out-of-sample period.
    \item Risks: Monthly volatility, Maximum Drawdown (MDD) and Downside Deviation.
    \item Financial Performance: Sharpe $\Bigl(\frac{\mathbb{E}[\textrm{Returns}]}{\textrm{Volatility}}\Bigr)$, Sortino $\Bigl(\frac{\mathbb{E}[\textrm{Returns}]}{\textrm{MDD}}\Bigr)$ and Calmar $\Bigl(\frac{\mathbb{E}[\textrm{Returns}]}{\textrm{Downside Deviation}}\Bigr)$ ratios are used as a gauge to measure risk-adjusted performance. We also include the average profit divided by the average loss $\Bigl(\frac{\textrm{Avg. Profits}}{\textrm{Avg. Loss}}\Bigr)$.
    \item Ranking Performance: Kendall's Tau, Normalised Discounted Cumulative Gain at $k$ (NDCG@$k$) \cite{jarvelinIREvaluationMethods2000}, which is suited for non-binary relevance (scoring) measures while also emphasising top returned results \cite{wuAdaptingBoostingInformation2010}. We note that $k$ is a pre-defined threshold, which we set fix at $k=100$ in our paper to cover the size of each of our long/short portfolios.
\end{enumerate}

\subsection{Results and Discussion}

\begin{figure}[ht!]
    \centering
    \caption{Cumulative Returns - Rescaled to Target Volatility.}
    \includegraphics[width=1\linewidth]{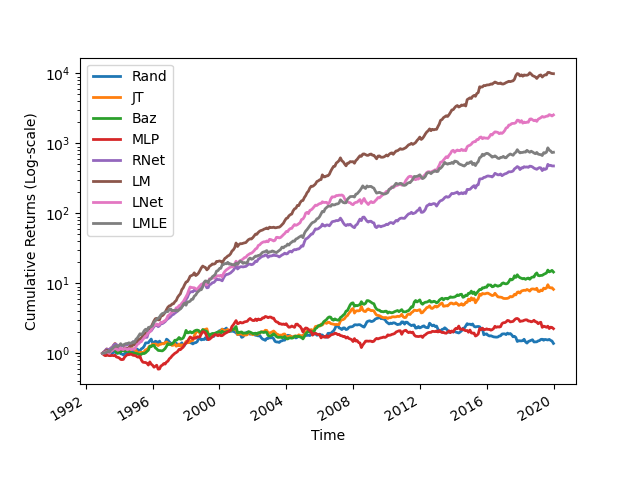}
    \label{fig:perf}
\end{figure}
\begin{table*}[h!]\centering
\caption{Performance Metrics -- Rescaled to Target Volatility.}
\label{table:fin_perf}
\ra{1.}\begin{tabular}{@{}rrrrrcrrrr@{}}
\toprule
& \multicolumn{4}{c}{\textbf{Benchmarks}}& \phantom{abc}& \multicolumn{4}{c}{\textbf{Learning to Rank Models}} \\
\cmidrule{2-5}\cmidrule{7-10}& Rand & JT & Baz & MLP && RNet & LM & LNet & LMLE\\
\midrule
E[returns]          & 0.024 & 0.092 & 0.112 & 0.044 && 0.243 & \textbf{*0.359} & 0.306 & 0.260 \\
Volatility          & 0.156 & 0.167 & 0.161 & 0.165 && 0.162 & 0.166 & \textbf{*0.155} & 0.162 \\
Sharpe              & 0.155 & 0.551 & 0.696 & 0.265 && 1.502 & \textbf{*2.156} & 1.970 & 1.611 \\
Downside Dev.       & 0.106 & 0.106 & 0.097 & 0.112 && 0.081 & \textbf{*0.067} & 0.068 & 0.071 \\
MDD                 & 0.584 & 0.328 & 0.337 & 0.641 && 0.294 & \textbf{*0.231} & 0.274 & 0.236 \\
Sortino             & 0.228 & 0.872 & 1.157 & 0.389 && 3.012 & \textbf{*5.321} & 4.470 & 3.647 \\
Calmar              & 0.042 & 0.281 & 0.333 & 0.068 && 0.828 & \textbf{*1.555} & 1.115 & 1.102 \\
\% +ve Returns      & 0.545 & 0.582 & 0.591 & 0.551 && 0.693 & \textbf{*0.762} & 0.715 & 0.681 \\
Avg. P / Avg. L     & 0.947 & 1.114 & 1.184 & 1.001 && 1.407 & 1.594 & \textbf{*1.679} & 1.534 \\
\bottomrule\end{tabular}
\end{table*}
\begin{table*}[h!]\centering
\caption{Ranking Metrics -- Average over all Rebalancing Months.}
\label{table:rank_perf}
\ra{1.}\begin{tabular}{@{}rrrrrcrrrr@{}}
\toprule
& \multicolumn{4}{c}{\textbf{Benchmarks}}& \phantom{abc}& \multicolumn{4}{c}{\textbf{Learning to Rank Models}} \\
\cmidrule{2-5}\cmidrule{7-10}& Rand & JT & Baz & MLP && RNet & LM & LNet & LMLE\\
\midrule
Kendall's Tau  & 0.000 & 0.016 & 0.013 & 0.008 && 0.032 & 0.032 & \textbf{*0.033} & 0.020 \\
NDCG@100 (Longs)  & 0.549 & 0.555 & 0.562 & 0.550 && 0.576 & 0.576 & \textbf{*0.578} & 0.565 \\
NDCG@100 (Shorts) & 0.552 & 0.562 & 0.555 & 0.564 && 0.575 & \textbf{*0.585} & 0.579 & 0.567 \\
\bottomrule\end{tabular}
\end{table*}
\begin{table*}[t!]
\caption{Performance Metrics – Decile Portfolios Rescaled to Target Volatility}
\label{table:deciles}
\centering\ra{1.05}\begin{tabular}{@{}rrrrrrrrrrrcr@{}}
\toprule
& \multicolumn{10}{c}{\textbf{Decile}} \\
\cmidrule{2-11}& 1 & 2 & 3 & 4 & 5 & 6 & 7 & 8 & 9 & 10 && $L-S$\\
\midrule
\textbf{Rand} \\
E[returns]      & 0.110 & 0.120 & 0.118 & 0.124 & 0.114 & 0.123 & 0.126 & 0.115 & 0.126 & 0.115 && 0.028 \\
Volatility      & 0.162 & 0.163 & 0.164 & 0.162 & 0.164 & 0.162 & 0.163 & 0.161 & 0.163 & 0.164 && 0.156 \\
Sharpe          & 0.675 & 0.737 & 0.721 & 0.769 & 0.697 & 0.763 & 0.772 & 0.710 & 0.772 & 0.702 && 0.177 \\
\textbf{JT} \\
E[returns]      & 0.059 & 0.071 & 0.096 & 0.108 & 0.122 & 0.127 & 0.137 & 0.151 & 0.146 & 0.146 && 0.094 \\
Volatility      & 0.165 & 0.164 & 0.165 & 0.164 & 0.163 & 0.163 & 0.163 & 0.162 & 0.160 & 0.156 && 0.167 \\
Sharpe          & 0.360 & 0.435 & 0.581 & 0.661 & 0.746 & 0.780 & 0.840 & 0.928 & 0.910 & 0.938 && 0.565 \\
\textbf{Baz} \\
E[returns]      & 0.095 & 0.094 & 0.083 & 0.094 & 0.092 & 0.109 & 0.124 & 0.137 & 0.160 & 0.185 && 0.107 \\
Volatility      & 0.163 & 0.163 & 0.162 & 0.162 & 0.163 & 0.162 & 0.162 & 0.162 & 0.163 & 0.163 && 0.161 \\
Sharpe          & 0.582 & 0.573 & 0.510 & 0.579 & 0.566 & 0.671 & 0.765 & 0.849 & 0.984 & 1.130 && 0.664 \\
\textbf{MLP} \\
E[returns]      & 0.072 & 0.112 & 0.122 & 0.114 & 0.127 & 0.126 & 0.130 & 0.135 & 0.124 & 0.132 && 0.097 \\
Volatility      & 0.163 & 0.161 & 0.163 & 0.162 & 0.163 & 0.162 & 0.163 & 0.163 & 0.163 & 0.164 && 0.168 \\
Sharpe          & 0.443 & 0.697 & 0.751 & 0.703 & 0.780 & 0.779 & 0.800 & 0.825 & 0.758 & 0.806 && 0.578 \\
\textbf{RNet} \\
E[returns]      & 0.043 & 0.067 & 0.079 & 0.095 & 0.115 & 0.121 & 0.140 & 0.147 & 0.163 & 0.202 && 0.246 \\
Volatility      & 0.164 & 0.164 & 0.164 & 0.164 & 0.164 & 0.163 & 0.161 & 0.162 & 0.161 & 0.163 && 0.161 \\
Sharpe          & 0.263 & 0.405 & 0.480 & 0.580 & 0.698 & 0.742 & 0.870 & 0.906 & 1.014 & 1.238 && 1.527 \\
\textbf{LM} \\
E[returns]      & 0.012 & 0.074 & 0.097 & 0.098 & 0.117 & 0.132 & 0.131 & 0.141 & 0.171 & 0.201 && 0.349 \\
Volatility      & 0.164 & 0.164 & 0.162 & 0.162 & 0.164 & 0.164 & 0.164 & 0.162 & 0.162 & 0.163 && 0.165 \\
Sharpe          & 0.075 & 0.449 & 0.599 & 0.606 & 0.716 & 0.806 & 0.800 & 0.868 & 1.053 & 1.232 && 2.107 \\
\textbf{LNet} \\
E[returns]      & 0.037 & 0.069 & 0.089 & 0.102 & 0.116 & 0.117 & 0.137 & 0.152 & 0.151 & 0.186 && 0.296 \\
Volatility      & 0.161 & 0.165 & 0.162 & 0.163 & 0.164 & 0.162 & 0.164 & 0.162 & 0.163 & 0.162 && 0.155 \\
Sharpe          & 0.232 & 0.416 & 0.549 & 0.628 & 0.711 & 0.720 & 0.837 & 0.938 & 0.929 & 1.148 && 1.911 \\
\textbf{LMLE} \\
E[returns]      & 0.059 & 0.080 & 0.088 & 0.110 & 0.109 & 0.123 & 0.126 & 0.151 & 0.163 & 0.193 && 0.244 \\
Volatility      & 0.165 & 0.164 & 0.164 & 0.165 & 0.161 & 0.164 & 0.162 & 0.161 & 0.160 & 0.163 && 0.160 \\
Sharpe          & 0.360 & 0.489 & 0.537 & 0.671 & 0.677 & 0.750 & 0.782 & 0.937 & 1.016 & 1.186 && 1.530 \\
\bottomrule\end{tabular}\end{table*}

To study the out-of-sample performance across various strategies, we chart their cumulative returns in Exhibit \ref{fig:perf} and tabulate key measures of financial performance in Exhibit \ref{table:fin_perf}. To allow for better comparability between strategy performance, we also apply an additional layer of volatility scaling at the portfolio level, bringing overall returns for each strategy in line with our $15\%$ target. All returns in this section are computed without transaction costs to focus on the raw predictive ability of the models. From both the plot and statistics, it is evident that our proposed class of LTR algorithms out-performs the set of benchmarks on all measures of performance -- with LambdaMART placed at the top for most metrics. 

In terms of profitability, the rankers significantly improve expected returns and the percentage win rate. The 'worst' LTR model significantly outperform the best reference benchmark for each metric considered.
While all models have been rescaled to trade around similar levels of volatility, LTR based strategies come across as being less subjected to huge drawdowns and downside risks.
On a performance basis, there is again an identical pattern of the lowest ranker dominating the best benchmark, and the best LTR model demonstrating substantial gains across various performance-based measures. This clear disparity in performance underscores the importance of learning the cross-sectional rankings as it leads to better performance for the momentum strategy. 
Further analysing the relative performance of models within each group, we first note that there is no clear superiority of the listwise LTR algorithms over their pairwise counterparts. One might have assumed that the listwise methods emerge more performant since they learn the broader listwise structure, which the results show is not necessarily the case. 
This might be explained by the inherently poor signal-to-noise ratio typical of financial datasets, which is further exacerbated by the limited size of the data used  -- specifically, the listwise approaches use approximately $12 \times N_{avg}$ samples per year while the pairwise methods have access to $12 \times N_{avg}^2$, where $N_{avg}$ is the average size of the cross-sectional universe at each month.
Across benchmarks, the Random model performed the worst as expected while the results of MLP are only marginally better. We suspect that this might also be the consequence of working with limited and noisy data which leads to overfitting, as well as the sub-optimality of the regress-then-rank approach utilised by MLP. Furthermore, the computed scores of MLP are essentially forecasts of (monthly) returns, which is regarded as a challenging problem \cite{naccaratoMarkowitzPortfolioOptimization2019} that is made even more so when all models used in this work are restricted to only using price-based data (See Section \ref{sec:pe_data_overview}).

By measuring an item's quality using \textit{graded} relevance and applying a discount over weights, the NDCG is well suited for assessing the quality of top-ranked items \cite{wangTheoreticalAnalysisNDCG2013} and is thus a widely used metric in the search literature \cite{liLearningRankInformation2011}.
The NDCG is particularly appropriate for the CSM strategy since it determines the extent to which models are able to accurately rank stocks by profitability -- a model that is able to rank in a more precise manner makes it likelier that top-ranked assets get selected for inclusion in the respective long/short portfolio. 
Given this, we assess all models based on the NDCG and set the cutoff $k=100$ to match the size of each of our long/short\footnote{To compute NDCG for shorts, we reversed our relevance scores which allows the most negative returns to attain the highest scores.} portfolios. 
From the set of compiled ranking metrics that is averaged across all months in Exhibit \ref{table:rank_perf}, all LTR models surpass the benchmarks when measured using NDCG@100 -- highlighting their ability to produce rankings that are more accurate, leading to better out-of-sample portfolio performance. With Kendall's Tau (rank correlation coefficient), we also see the same pattern of out-performance, noting that this time ranking quality is assessed across the \textit{entire} list of assets. 

\begin{figure*}
    \centering
    \caption{Cumulative Returns - Decile Portfolios Rescaled to Target Volatility}
    \includegraphics[width=1\linewidth]{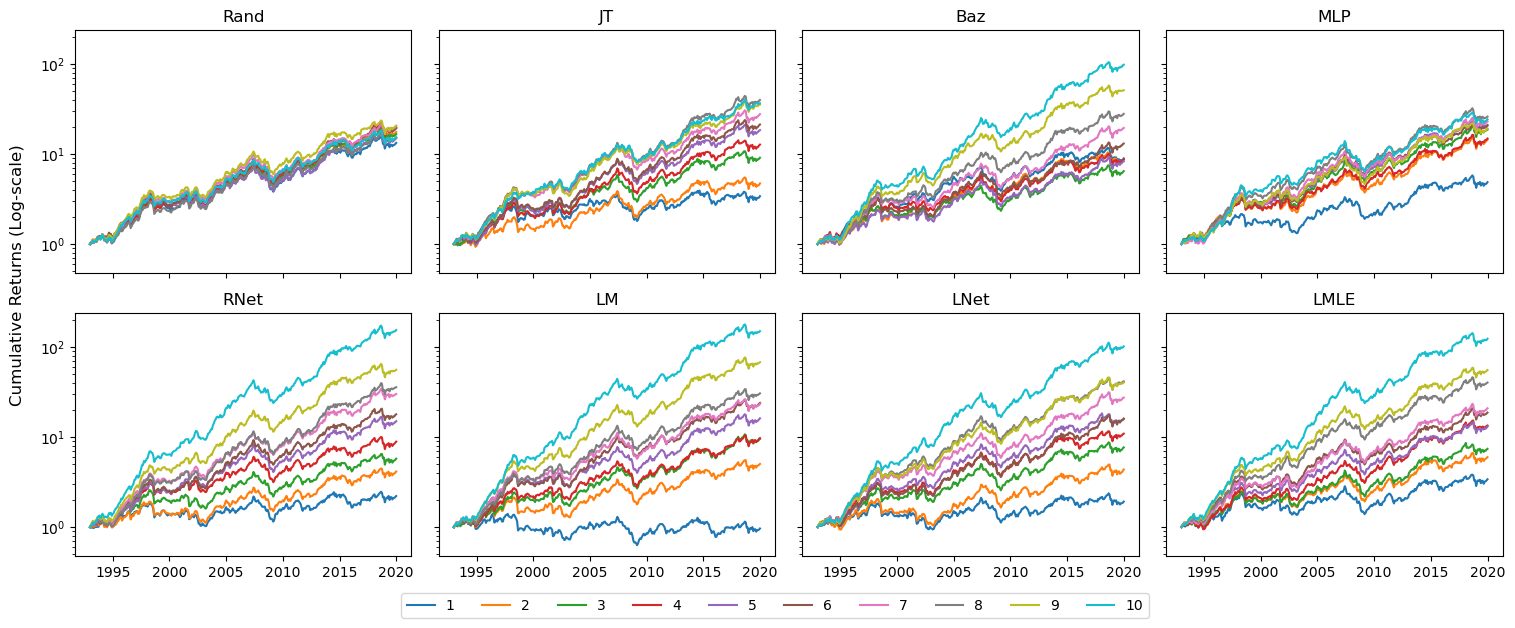}
    \label{fig:witch}
\end{figure*}

To further examine how rankings quality is linked to out-of-sample results, we construct long-only decile portfolios: 
At each month, these are formed by partitioning the asset universe into equally weighted deciles based on the signals/scores produced by their respective models. For instance, assets in the top (long) decile for MLP would contain the highest $10\%$\footnote{This differs from our earlier approach of using 100 instruments for each long and short portfolios at all times.} of the model's predictions.  Returns are computed similarly to Equation \eqref{eqn:csm_rets} but with a decile membership indicator $D^{(i)}_{\tau_m}$ used in place of $X^{(i)}_{\tau_m}$:
\begin{align}
r^{DEC}_{\tau_m,\tau_{m+1}} = \frac{1}{n_{\tau_m}} \sum^{n_{\tau_m}}_{i=1} D^{(i)}_{\tau_m} \frac{\sigma_{tgt}}{\sigma^{(i)}_{\tau_m}} r^{(i)}_{\tau_m,\tau_{m+1}}
\label{eqn:csm_dec_rets} 
\end{align}
where $D^{(i)}_{\tau_m} = \{0, 1\}$ and takes the value of $1$ for the decile of interest and $0$ otherwise. We also perform an additional level of scaling at the portfolio level. 
Referring to the summary results (Exhibit \ref{table:deciles}), there is a general trend of returns and Sharpe increasing from decile 1 to 10 across all strategies except Random which emphasises the consistency of the momentum factor. 
More important is the steeper rise of these figures for the LTR models stemming from their ability to place
assets in their appropriate deciles with a greater degree of precision -- leading to a greater difference in returns between the decile 1 and decile 10 portfolios.
The plot of decile portfolio returns across strategies (Exhibit \ref{fig:witch}) cement this observation, illustrating the connection between the model's ranking ability and the dispersion across return streams. 
This relationship is most pronounced for the group of LTR models which echoes the preceding statistics, thus validating our hypothesis that better asset rankings improves strategy performance.

\section{Conclusions}
Focusing on cross-sectional momentum as a demonstrative use-case, we introduce Learning to Rank algorithms as a novel way of ranking assets which is an important step required by cross-sectional systematic strategies. Additionally, the modular framework underpinning these algorithms allows additional feature inputs to be flexibly incorporated -- providing a generalisable platform for a broader set of cross-sectional strategies.
In learning the relational (pairwise or listwise) structure across assets that both heuristics and regress-then-rank approaches only superficially capture, we obtain a more accurate ranking across instruments. This translates to Sharpe Ratios being boosted approximately threefold over traditional approaches, as well providing clear and significant improvements to both performance and ranking-quality related measures.

Some directions for future work include innovating in terms of architecture or model ensembling to further improve strategy performance, as well as studying the effectiveness of these ranking techniques on higher frequency data (e.g. order-book) and asset classes. 

\bibliography{references}
\bibliographystyle{IEEEtran}


\section{Appendix}
\subsection{Learning to Rank For Cross-sectional Momentum}\label{sec:app_ltr4csm}

Learning to rank (LTR) is a supervised learning task involving training and testing phases. Document retrieval is the standard problem setting to make concrete the LTR framework, and we follow this convention. For training, we are provided with a set of queries $\bm{\mathcal{Q}}=\{q_1, ..., q_m\}$. Each query $q_i$ has an associated list of documents $\bm{d}_i=\{d_i^{(1)}, ... , d_i^{(n_i)}\}$ where $n_i$ is the total number of documents for $q_i$, and an accompanying set of document labels $\bm{y}_i=\{y_i^{(1)}, ... , y_i^{(n_i)}\}$ where the labels represent grades. 
Letting $\bm{\mathcal{Y}}=\{\mathcal{Y}_1, ..., \mathcal{Y}_\ell\}$ be the label set, we have $y_i^{(j)}\in \bm{\mathcal{Y}} \textrm { for } \forall \textrm{ } i, j$. We also have $\mathcal{Y}_\ell \succ \mathcal{Y}_{\ell-1} \succ ... \succ \mathcal{Y}_1$ where $\succ$ stands for the order relation -- a higher grade on a given document implies a stronger relevance of the document with respect to its query. For each query-document pair, a feature vector $x_i^{(j)}=\phi\bigl(q_i, d_i^{(j)}\bigr)$ can be formed, noting that $\phi(\cdot)$ is a feature function, $i\in\{1, ... , m\}$ and $j\in\{1, ... , n_i\}$. Letting $\bm{x}_i=\{x^{(1)}, ... , x^{(n_i)}\}$, we can assemble the training set $\bigl\{\bm{x}_i, \bm{y}_i\bigr\}^m_{i=1}$. 
The goal of LTR is to learn a function $f$ that predicts a score $f(x^{(i)}_{m+1})=z^{(i)}_{m+1}$ when presented with an out-of-sample input $x^{(i)}_{m+1}$. For more details, we point the reader to \cite{liLearningRankInformation2011}.

\begin{figure}
    \centering
    \caption{Learning to Rank for Cross-Sectional Momentum}
    \includegraphics[width=1\linewidth]{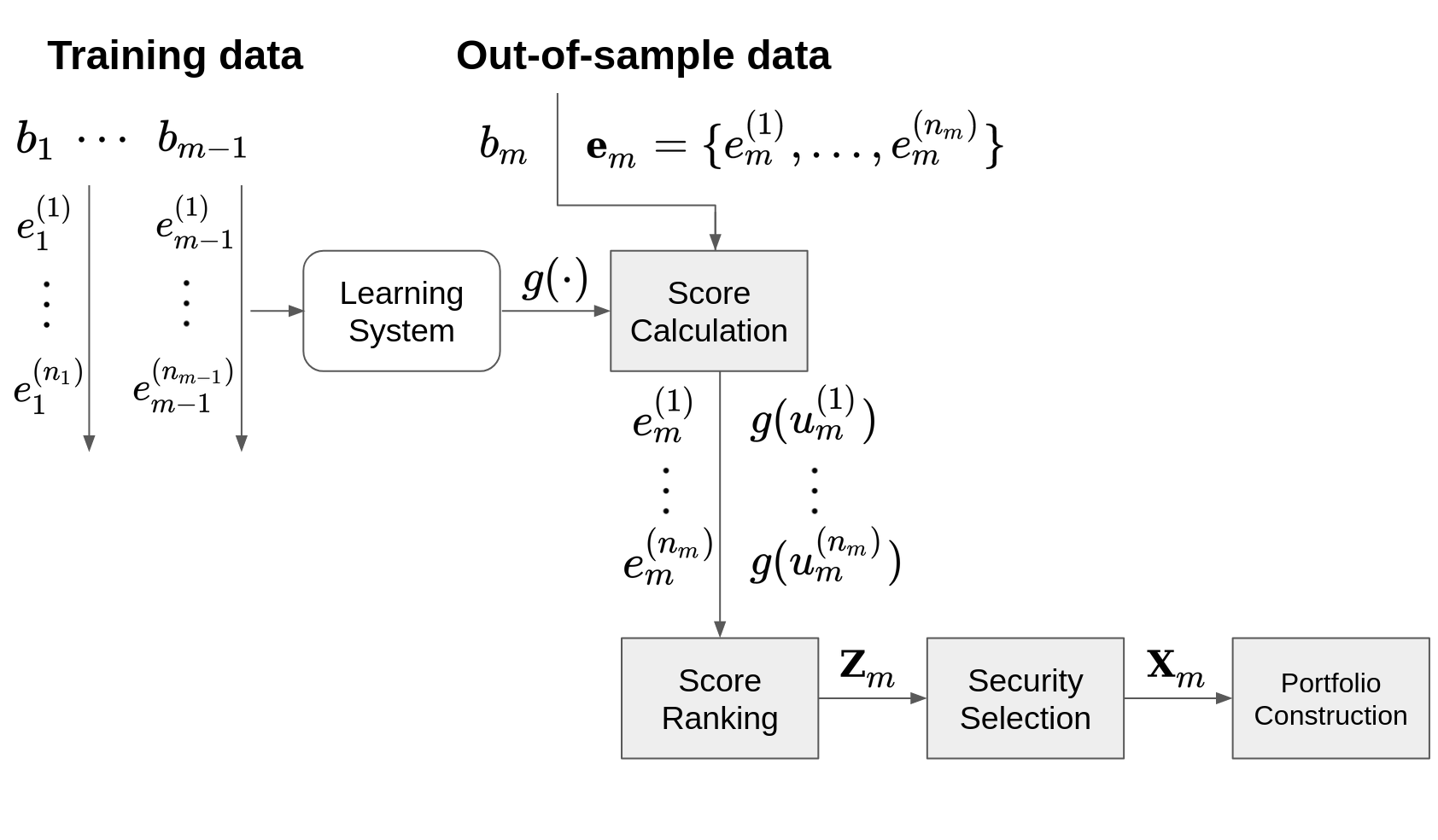}
    \label{fig:ltr_csm}
\end{figure}

Transposing the preceding framework to the momentum strategy, we can treat each query as being analogous to a portfolio rebalancing event, while an associated document and its accompanying label can be respectively thought of as an asset and its assigned decile at the \textit{next} rebalance based on some performance measure (conventionally taken to be returns). Exhibit \ref{fig:ltr_csm} provides a schematic of this adaptation which we further make concrete.
For training, let $\bm{\mathcal{B}}=\{b_1, ..., b_{m-1} \}$ be a series of monthly rebalances where at each $b_i$ we have a set of equity instruments $\bm{e}_i=\{ e^{(1)}_i, ... , e^{(n_i)}_i \}$ and the set of assigned deciles  $\bm{\delta}_{i+1}= \{  \delta^{(1)}_{i+1}, ... , \delta^{(n_i)}_{i+1} \}$ where $\delta^{(j)}_{i+1} \in \bm{\mathcal{D}} = \{\mathcal{D}_1, ..., \mathcal{D}_{10} \} \textrm{ for } \forall \textrm{ }i, j$. Similar to above, $k > l \Rightarrow \mathcal{D}_k \succ \mathcal{D}_l \textrm{ for } k, l \in \{1, ..., 10\}$. With each rebalance-asset pair, we can form the feature vector $u^{(j)}_i = \phi(b_i, e^{(j)}_i)$ as well as the broader training set $\bigl\{\bm{u}_i, \bm{\delta}_{i+1}\bigr\}^{m-1}_{i=1}$ where  $\bm{u}_i=\{u^{(1)}_i, ..., u^{(n_i)}_i \}$. Note that other features can be incorporated for $u^{(j)}_i$ -- allowing different types of cross-sectional strategies to be developed. Presented with set of feature vectors for testing at interval $m$ with a trained function $g$ produced by the Learning System, we compute the set of scores $g(\bm{u}_m)$ at the Score Calculation phase and then go on to form the long/short portfolios by following the instructions in Section \ref{sec:csm_framework}.

\subsection{Additional Training Details}\label{app:training_details}

\textit{Python Libraries:} LambdaMART uses {\tt XGBoost} \cite{chenXGBoostScalableTree2016}, while the others -- RankNet, ListNet, ListMLE are developed using {\tt Tensorflow} \cite{abadiTensorFlowLargescaleMachine2015}. 

\textit{Hyperparameter Optimisation:} Hyperparameters assume discrete values and are tuned using {\tt HyperOpt} \cite{bergstraHyperoptPythonLibrary2015}. For LambdaMART, we refer to the hyperparameters as they are named in the {\tt XGBoost} library.

\vspace{10pt}
\textbf{Multi-layer Perceptron (MLP)}
\begin{itemize}
    \item Dropout Rate -- [0.0, 0.2, 0.4, 0.6, 0.8]
    \item Hidden Width -- [64, 128, 256, 512, 1024, 2048]
    \item Max Gradient Norm -- $[10^{-3}, 10^{-2}, 10^{-1}, 1, 10]$
    \item Learning Rate -- $[10^{-6}, 10^{-5}, 10^{-4}, 10^{-3}, 10^{-2},\\ 10^{-1}, 1]$
    \item Minibatch Size -- [64, 128, 256, 512 1024]
\end{itemize}

\vspace{10pt}
\textbf{RankNet}
\begin{itemize}
    \item Dropout Rate -- [0.0, 0.2, 0.4, 0.6, 0.8]
    \item Hidden Width -- [64, 128, 256, 512, 1024, 2048]
    \item Max Gradient Norm -- $[10^{-3}, 10^{-2}, 10^{-1}, 1, 10]$
    \item Learning Rate -- $[10^{-6}, 10^{-5}, 10^{-4}, 10^{-3}, 10^{-2},\\ 10^{-1}, 1]$
    \item Securities used to form Pairs for Minibatch -- [64, 128, 256, 512 1024]
\end{itemize}

\vspace{10pt}
\textbf{LambdaMART}
\begin{itemize}
    \item `objective' -- `rank:pairwise'
    \item `eval\_metric' -- `ndcg'
    \item `eta' -- $[10^{-6}, 10^{-5}, 10^{-4}, 10^{-3}, 10^{-2}, 10^{-1}, 1]$
    \item `num\_boost\_round' -- $[5, 10, 20, 40, 80, 160, 320]$
    \item `max\_depth' -- $[2, 4, 6, 8, 10]$
    \item `tree\_method' -- `gpu\_hist'
\end{itemize}

\vspace{10pt}
\textbf{ListMLE, ListNet}
\begin{itemize}
    \item Dropout Rate -- [0.0, 0.2, 0.4, 0.6, 0.8]
    \item Hidden Width -- [64, 128, 256, 512, 1024, 2048]
    \item Max Gradient Norm -- $[10^{-3}, 10^{-2}, 10^{-1}, 1, 10]$
    \item Learning Rate -- $[10^{-8}, 10^{-7}, 10^{-6}, 10^{-5}, 10^{-4}]$
    \item Minibatch Size -- [1, 2, 4, 8, 16]
\end{itemize}

\end{document}